\documentclass[prb,aps,twocolumn,amsmath,amssymb]{revtex4}
\usepackage{graphicx,bm}
\usepackage{dcolumn}

\begin{document}

\title{Number-Phase Fluctuations in Isolated Superconductors}

\author{Xiaotian Si, Wataru Kohno, and Takafumi Kita}
\affiliation{Department of Physics, Hokkaido University, Sapporo 060-0810, Japan}%

\begin{abstract}
We improve the Bardeen-Cooper-Schrieffer wave function with a fixed particle number so as to incorporate
many-body correlations beyond the mean-field treatment.
It is shown that the correlations lower the ground-state energy far more than Cooper-pair condensation in the weak-coupling region.
Moreover, they naturally bring a superposition over the number of condensed particles. 
Thus, Cooper-pair condensation is special among the various bound-state formations of quantum mechanics
in that number fluctuations are necessarily present in the condensate through the dynamical exchange of particles 
with the non-condensate reservoir.
On the basis of this result, we propose $\varDelta N_{\rm con}\cdot \varDelta \phi\gtrsim 1$ as the uncertainty relation relevant
to the number-phase fluctuations in superconductors and superfluids,
where the number of condensed particles $N_{\rm con}$ is used instead of the total particle number $N$. 
The formula implies that a macroscopic phase $\phi$ can be established even in number-fixed superconductors and superfluids 
since $\varDelta N_{\rm con}\gg 1$.
\end{abstract}

\maketitle

\section{Introduction}

One of the most controversial issues in the Bardeen-Cooper-Schrieffer (BCS) theory,\cite{BCS} which is remarkably successful in describing
weak-coupling superconductors, may be the superposition over the number of condensed particles
in their variational ground-state wave function. 
This is apparently incompatible with particle-number conservation, which manifestly holds in any closed system,
as noted by Schrieffer from the beginning \cite{Schrieffer11} and emphasized by Peierls \cite{Peierls} and Leggett.\cite{Leggett06}
On the other hand, the superposition was used by Anderson \cite{Anderson66} in the context of Bose-Einstein condensation
to discuss the emergence of a well-defined macroscopic phase, called 
{\em spontaneously broken gauge symmetry},  \cite{Anderson64,Leggett91}
as the key ingredient for superfluidity and the Josephson effect.
Thus, particle-number fluctuations seem indispensable for bringing macroscopic coherence to the system,
which were originally traced by Anderson to the exchange of particles between subsystems.  \cite{Anderson66}
However, a question may be raised regarding this identification because there are definitely no fluctuations in the total particle number of any closed system. \cite{Leggett91,Leggett06}
Are the fluctuations real or a mere artifact in the mathematical treatment of superconductivity?
If the former is the case, where do they originate from?
How can we define a macroscopic wave function with a well-defined phase in isolated superconductors?
We aim to answer these questions by improving the BCS wave function with a fixed particle number.

Weak-coupling superconductors have been described theoretically within the mean-field framework.
The corresponding ground state with $N$ fermions has been identified as the antisymmetrized product of $N/2$ Cooper pairs with no superposition,\cite{Leggett06,SchriefferText,Ambegaokar} which may thereby have no well-defined phase.\cite{Anderson66}
Now, we will see what happens to this wave function when we incorporate many-body correlations beyond the mean-field treatment.
Our physical motivation lies in the following observation: 
the pair condensation energy 
in the weak-coupling region is exponentially small, $\sim \exp(-1/g)$ with $g\!>\!0$ a dimensionless coupling constant, 
whereas the correlation energy is proportional to $g^2$ and also negative for any type of interaction, 
as seen by the second-order perturbation in terms of the interaction.
In other words, the correlations lower the ground-state energy far more than Cooper-pair condensation 
for $g\ll 1$.
This fact implies that, formally speaking, Cooper-pair condensation should be studied only after the correlation effects have been incorporated.
We incorporate the correlation effects to show explicitly that the correlations produce finite non-condensed particles in the ground state, which work as a particle reservoir for the condensate
to naturally yield the superposition,
in exactly the same way as in the case of interacting Bose-Einstein condensates.\cite{Kita17}
Thus, the superposition is a real physical entity that exists in any isolated superconductor or superfluid. 
Note in this context that the superposition and coherence have so far been discussed mostly in terms of condensed particles alone.\cite{Anderson66,Leggett91,JY96,CD97}

This paper is organized as follows. Section 2 presents the formulation.
Section 3 gives numerical results. Section 4 presents concluding remarks.
Appendix A derives equations to minimize the variational ground-state energy in detail.
Appendix B describes how to perform triple sums over wave vectors efficiently in the numerical calculations.

\section{Formulation}

\subsection{Model}
To make our problem mathematically well-defined and tractable, we consider a simplified model 
that consists of $N$ identical fermions 
($N$: even) with mass $m$ and spin $\frac{1}{2}$ interacting via a two-body attractive potential $U(r)$ in a box of volume $V$.\cite{Leggett80,NSR85}
The Hamiltonian is given explicitly by
\begin{align}
\hat{H}\equiv \sum_{{\bm k}\alpha}\varepsilon_{k}\hat{c}_{{\bm k}\alpha}^\dagger \hat{c}_{{\bm k}\alpha}+\frac{1}{2V}
\sum_{{\bm k}{\bm k}'{\bm q}}\sum_{\alpha\alpha'}U_q \hat{c}_{{\bm k}+{\bm q}\alpha}^\dagger \hat{c}_{{\bm k}-{\bm q}\alpha'}^\dagger  \hat{c}_{{\bm k}'\alpha'}
\hat{c}_{{\bm k}\alpha} ,
\label{hatH}
\end{align}
where $\varepsilon_{k}$ and $U_q$ are
\begin{align}
\varepsilon_{k}\!\equiv\! \frac{\hbar^2k^2}{2m},\hspace{7mm}
U_q\!\equiv\! \int U(r)\,e^{-i{\bm q}\cdot{\bm r}}d^3 r,
\end{align}
and operators $(\hat c_{{\bm k}\alpha},\hat c_{{\bm k}\alpha}^\dagger)$ satisfy 
the anticommutation relations of fermions with $\alpha=\uparrow,\downarrow$ for $\alpha=\frac{1}{2},-\frac{1}{2}$, respectively.

\subsection{Number-fixed BCS wave function}

Anticipating condensation into a homogeneous $s$-wave pairing for this model, 
we introduce the pair creation operator by
\begin{align}
\hat \pi^\dagger \equiv \sum_{{\bm k}}\phi_{{\bm k}} \hat{c}_{{\bm k}\uparrow}^\dagger\hat{c}_{-{\bm k}\downarrow}^\dagger, 
\end{align}
with $\phi_{\bm k}$ denoting the Fourier coefficient of the bound-state wave function describing a single Cooper pair. 
The number-fixed BCS wave function is given in terms of $\hat\pi^\dagger$ by \cite{Leggett06,SchriefferText,Ambegaokar,Leggett80,KitaText}
\begin{align}
|\Phi_{N}^{\rm BCS}\rangle \equiv {\cal A}_{N/2}^{-1/2} \frac{\displaystyle \left(\hat\pi^\dagger\right)^{N/2}}{\displaystyle (N/2)!}|0\rangle ,
\label{|Phi>}
\end{align}
where ${\cal A}_{n}\!\equiv\! \langle0|\hat\pi^{n}(\hat\pi^\dagger)^{n}|0\rangle/(n!)^2$ normalizes the ket and $|0\rangle$ is defined by
$\hat{c}_{{\bm k}\alpha}|0\rangle\!=\!0$.
Equation (\ref{|Phi>}) is the $N$-particle projection of the original BCS wave function $|\Phi^{\rm BCS}\rangle \!\propto\! \exp(\hat\pi^\dagger)|0\rangle$. \cite{BCS,Leggett06,SchriefferText,KitaText}
It satisfies 
\begin{align}
\hat\gamma_{{\bm k}\alpha}|\Phi_{N}^{\rm BCS}\rangle\!=\!0,
\label{gamma|Phi_BCS>=0}
\end{align}
i.e., the ket is characterized as the vacuum of the number-conserving Bogoliubov operator \cite{Ambegaokar}
\begin{align}
\hat\gamma_{{\bm k}\alpha}\equiv u_{{\bm k}}\hat c_{{\bm k}\alpha}-(-1)^{\frac{1}{2}-\alpha}\,v_{{\bm k}}\hat c_{-{\bm k}-\alpha}^\dagger\hat{P} ,
\label{gamma-c}
\end{align}
where $(u_{{\bm k}},v_{{\bm k}})$ denote 
\begin{align}
u_{{\bm k}}\equiv (1+|\phi_{\bm k}|^2)^{-1/2},\hspace{7mm}v_{{\bm k}}\equiv u_{{\bm k}}\phi_{{\bm k}},
\label{uv}
\end{align}
satisfying $u_{{\bm k}}^2+|v_{{\bm k}}|^2=1$, and operators $\hat{P}$ and $\hat{P}^\dagger$ are defined by
\begin{subequations}
\label{P-def}
\begin{align}
\hat{P} |\Phi_{N}^{\rm BCS}\rangle =  |\Phi_{N-2}^{\rm BCS}\rangle,\hspace{7mm}
\hat{P}^\dagger |\Phi_{N}^{\rm BCS}\rangle =  |\Phi_{N+2}^{\rm BCS}\rangle.
\end{align}
Thus, $\hat{P}$ $(\hat{P}^\dagger)$ decreases (increases) the number of Cooper pairs by one.
They satisfy 
\begin{align}
(\hat{P}^\dagger)^\nu\hat{P}^\nu=\hat{P}^\nu(\hat{P}^\dagger)^\nu=1
\end{align}
\end{subequations}
asymptotically for $\nu\ll N/2$ and 
can be treated as commutative with $(\hat c_{{\bm k}\alpha},\hat c_{{\bm k}\alpha}^\dagger)$.  \cite{Ambegaokar}
One can thereby show that $\hat\gamma_{{\bm k}\alpha}$ also obeys the anticommutation relations of fermions.

\subsection{Improved wave function with correlations}

\begin{figure}[t]
\begin{center}
\includegraphics[width=0.8\linewidth]{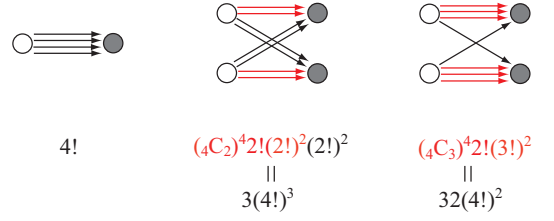}
\caption{(Color online) Diagrammatic expansion of $\ln{\cal B}_N$ up to the second order in $\hat{\pi}_4$. 
An open (filled) circle with four outgoing (incoming) arrows denotes $\hat{\pi}_4^\dagger$ ($\hat{\pi}_4$).
The weight below each figure represents the number of combinations that realize the connection.
\label{Fig1}}
\end{center}
\end{figure}

Now, we incorporate many-body correlations into Eq.\ (\ref{|Phi>}).
Equation (\ref{gamma|Phi_BCS>=0}) indicates
that the Bogoliubov quasiparticles are absent from the mean-field BCS ground state given by Eq.\ (\ref{|Phi>}).
With this observation, we investigate the possibility that some of the quasiparticle states become occupied 
owing to many-body correlations. 
To this end, we introduce the number-conserving correlation operator
\begin{align}
\hat\pi_4^\dagger\equiv &\,\frac{1}{4!}\sum_{\kappa_1}\sum_{\kappa_2}\sum_{\kappa_3}\sum_{\kappa_4}w_{\kappa_1\kappa_2\kappa_3\kappa_4}
\hat\gamma_{\kappa_1}^\dagger\hat\gamma_{\kappa_2}^\dagger\hat\gamma_{\kappa_3}^\dagger\hat\gamma_{\kappa_4}^\dagger \hat{P}^2 ,
\label{pi_4}
\end{align}
where $\kappa_j$ denotes $\kappa_j\equiv {\bm k}_j\alpha_j$,
and $w_{\kappa_1\kappa_2\kappa_3\kappa_4}$ is a variational parameter that is antisymmetric with respect to any permutation of $(\kappa_1,\kappa_2,\kappa_3,\kappa_4)$ by definition. 
This operator $\hat\pi_4^\dagger$ describes the process where two Cooper pairs are broken up into four quasiparticles.
Our variational wave function is given in terms of Eqs.\ (\ref{|Phi>}) and (\ref{pi_4}) by
\begin{align}
|\Phi_N\rangle\equiv {\cal B}_N^{-1/2}\exp(\hat\pi_4^\dagger)|\Phi_{N}^{\rm BCS}\rangle ,
\label{|Psi>}
\end{align}
where ${\cal B}_N$ denotes the normalization constant.
This $|\Phi_N\rangle$ indeed has finite occupations of quasiparticles when $w_{\kappa_1\kappa_2\kappa_3\kappa_4}\neq 0$
is realized.
It should also be noted that operating the exponential function $\exp(\hat\pi_4^\dagger)$ on $|\Phi_{N}^{\rm BCS}\rangle$,
among other possible functions of $\hat\pi_4^\dagger$, has a technical advantage that
we can use linked cluster expansions in the evaluation of various physical quantities. 
For example,  the normalization constant ${\cal B}_N$ is obtained as
\begin{align}
{\cal B}_N\equiv &\,\langle\Phi_{N}^{\rm BCS}|\exp(\hat\pi_4)\exp(\hat\pi_4^\dagger)|\Phi_{N}^{\rm BCS}\rangle 
\notag \\ 
=&\,\exp\left(\frac{1}{4!}\sum_{\kappa_1}\sum_{\kappa_2}\sum_{\kappa_3}\sum_{\kappa_4}|w_{\kappa_1\kappa_2\kappa_3\kappa_4}|^2+\cdots \right).
\label{calB_N}
\end{align}
The exponent in the second expression is expressible as Fig.\ \ref{Fig1} in terms of connected Feynman diagrams,\cite{Kita17}
and the first term denotes the lowest-order contribution;
we omit the higher-order terms in the present weak-coupling consideration. 

It will turn out below that Eqs.\ (\ref{gamma|Phi_BCS>=0}) and (\ref{calB_N}) suffice to perform an evaluation of the ground-state
energy up to the leading order in the correlation parameter $w_{\kappa_1\kappa_2\kappa_3\kappa_4}$ beyond the BCS theory.

\subsection{Expression for the ground-state energy}

Evaluation of the variational ground-state energy 
\begin{align}
{\cal E}\equiv \langle \Phi_N |\hat{H}|\Phi_N\rangle
\label{calE-def}
\end{align}
can be performed in exactly the same way as that for the 
interacting Bose-Einstein condensates. \cite{Kita17} 
Specifically, we express 
\begin{align}
\hat{c}_{{\bm k}\alpha}
= u_{{\bm k}}\hat\gamma_{{\bm k}\alpha}+(-1)^{\frac{1}{2}-\alpha}\,v_{{\bm k}}\hat\gamma_{-{\bm k}-\alpha}^\dagger\hat{P}
\end{align}
based on Eq.\ (\ref{gamma-c}), transform $\hat{H}$ into the normal order in $\hat\gamma_{{\bm k}\alpha}$,
and evaluate  ${\cal E}$ subsequently. 
A new ingredient here compared with the BCS theory is the finite average:
\begin{align}
\langle\Phi_N|\hat\gamma_{\kappa_1}^\dagger\hat\gamma_{\kappa_2}^\dagger\hat\gamma_{\kappa_3}^\dagger\hat\gamma_{\kappa_4}^\dagger \hat{P}^2|\Phi_N\rangle 
= \frac{\delta\ln {\cal B}_N}{\delta w_{\kappa_1\kappa_2\kappa_3\kappa_4}}\approx w_{\kappa_1\kappa_2\kappa_3\kappa_4}^*,
\label{gamma-4}
\end{align}
where we have used Eq.\ (\ref{calB_N}). 
Also noting 
\begin{align*}
\hat\gamma_{\kappa}|\Phi_N\rangle=&\,{\cal B}_N^{-1/2}[\hat\gamma_{\kappa},e^{\hat\pi_4^\dagger}]|\Phi_N^{\rm BCS}\rangle=[\hat\gamma_{\kappa},\hat\pi_4^\dagger] |\Phi_N\rangle
\\
= &\, \frac{1}{3!}\sum_{\kappa_2\kappa_3\kappa_4}w_{\kappa\kappa_2\kappa_3\kappa_4}
\hat\gamma_{\kappa_2}^\dagger\hat\gamma_{\kappa_3}^\dagger\hat\gamma_{\kappa_4}^\dagger|\Phi_N\rangle,
\end{align*}
we find another finite average
\begin{align}
\eta_{{\bm k}}\equiv \langle\Phi_N|\hat\gamma_{{\bm k}\alpha}^\dagger\hat\gamma_{{\bm k}\alpha}|\Phi_N\rangle \approx \frac{1}{3!} \sum_{\kappa_2\kappa_3\kappa_4}|w_{\kappa\kappa_2\kappa_3\kappa_4}|^2,
\label{eta_k}
\end{align}
where we have omitted the possibility of spin polarization; accordingly, 
 $|w_{\kappa\kappa_2\kappa_3\kappa_4}|^2$ in Eq.\ (\ref{eta_k}) should be interpreted as the average of $\kappa={\bm k}\uparrow$ and $\kappa={\bm k}\downarrow$; we also assume that $w_{\kappa_1\kappa_2\kappa_3\kappa_4}$ is real from now on.
 
It is convenient to introduce two basic expectations with $|\Phi_N\rangle$,
\begin{subequations}
\label{nF-def}
\begin{align}
n_{{\bm k}}\equiv &\,\langle\Phi_N|\hat{c}_{{\bm k}\alpha}^\dagger\hat{c}_{{\bm k}\alpha}|\Phi_N\rangle=
v_{{\bm k}}^2+(u_{{\bm k}}^2-v_{{\bm k}}^2)\eta_{{\bm k}} ,
\label{n_k}
\\
F_{{\bm k}}\equiv&\, \langle\Phi_N|\hat{P}^\dagger\hat{c}_{-{\bm k}\downarrow}\hat{c}_{{\bm k}\uparrow}|\Phi_N\rangle=
u_{{\bm k}}v_{{\bm k}}(1-2\eta_{{\bm k}}) ,
\label{F_k}
\end{align}
\end{subequations}
where we have assumed that $\phi_{{\bm k}}$ is also real.
Note that $F_{{\bm k}}=F_{{\bm k}}^*$ implies that $F_{{\bm k}}$ can also be written as 
$F_{{\bm k}}=\langle\Phi_N|\hat{c}_{{\bm k}\uparrow}^\dagger \hat{c}_{-{\bm k}\downarrow}^\dagger\hat{P}\,|\Phi_N\rangle$.
Using Eqs.\ (\ref{gamma-4}) and (\ref{nF-def}), we can concisely express Eq.\ (\ref{calE-def}) in the weak-coupling region as
\begin{align}
{\cal E}=&\, 2\sum_{{\bm k}}\varepsilon_kn_{{\bm k}}+
 \frac{1}{V}\sum_{{\bm k}{\bm k}'}(2U_0 -U_{|{\bm k}-{\bm k}'|}) n_{{\bm k}}n_{{\bm k}'} 
\notag \\
&\,+\frac{1}{V}\sum_{{\bm k}{\bm k}'}U_{|{\bm k}-{\bm k}'|}F_{{\bm k}}F_{{\bm k}'}
\notag \\
&\, +\frac{1}{V}\sum_{{\bm k}_1{\bm k}_2{\bm k}_3{\bm k}_4}\delta_{{\bm k}_1+{\bm k}_2+{\bm k}_3+{\bm k}_4,{\bm 0}}\,U_{|{\bm k}_1+{\bm k}_3|}\,
u_{{\bm k}_1}u_{{\bm k}_2}v_{{\bm k}_3}v_{{\bm k}_4}
\notag \\
&\,\times
\sum_{\alpha\alpha'}(-1)^{1-\alpha-\alpha'}
 w_{{\bm k}_1\alpha{\bm k}_2\alpha'{\bm k}_3-\alpha'{\bm k}_4-\alpha} .
\label{calE}
\end{align}
The fourth term is the correlation energy characteristic of the present theory, 
whereas the first, second, and third terms are the kinetic, Hartree-Fock, and pair-condensation energies, respectively.
Setting $\eta_{{\bm k}}$ and $w_{\kappa_1\kappa_2\kappa_3\kappa_4}$ to zero in Eq.\ (\ref{calE}) reproduces the
BCS expression  for the ground-state energy including the Hartree-Fock contribution.

\subsection{Minimization of $\,{\cal E}$}

To minimize Eq.\ (\ref{calE}) for a fixed $N$, 
we incorporate the constraint 
\begin{align}
2\sum_{{\bm k}}n_{{\bm k}}=N
\label{N-constraint}
\end{align}
given in terms of Eq.\ (\ref{n_k}) by the method of Lagrange multipliers.
Specifically, we introduce the functional
\begin{align}
\bar{\cal E}\equiv {\cal E}-\mu\left(2\sum_{{\bm k}}n_{{\bm k}}-N\right) 
\label{barcalE}
\end{align}
with $\mu$ denoting the Lagrange multiplier,
and set its first variations with respect to $\phi_{{\bm k}}$ and $w_{\kappa_1\kappa_2\kappa_3\kappa_4}$ equal to zero simultaneously.
These variations can be calculated straightforwardly but rather tediously as detailed in Appendix A, 
which is outlined as follows.
The equation for $\delta w_{\kappa_1\kappa_2\kappa_3\kappa_4}$ turns out to be linear in 
$w_{\kappa_1\kappa_2\kappa_3\kappa_4}$ and can be solved explicitly.
Substitution of the resultant expression into the equation for $\delta\phi_{{\bm k}}$ yields
\begin{align}
\phi_{{\bm k}}=\frac{-\xi_{{\bm k}}+E_{{\bm k}}}{\Delta_{{\bm k}}},\hspace{7mm}E_{{\bm k}}\equiv \sqrt{\xi_{{\bm k}}^2+\Delta_{{\bm k}}^2},
\label{phi_k}
\end{align}
with which Eq.\ (\ref{uv}) acquires the standard BCS expression
\begin{align}
u_{{\bm k}}=\sqrt{\frac{1}{2}\left(1+\frac{\xi_{{\bm k}}}{E_{{\bm k}}}\right)}\,\,,\hspace{7mm}
v_{{\bm k}}=\sqrt{\frac{1}{2}\left(1-\frac{\xi_{{\bm k}}}{E_{{\bm k}}}\right)} \,\,.
\label{uv-sol}
\end{align}
However, correlations are now incorporated in the single-particle energy $\xi_{{\bm k}}$ and energy gap $\Delta_{{\bm k}}$ as
\begin{subequations}
\label{xiDelta-def}
\begin{align}
\xi_{{\bm k}}= &\, \varepsilon_k-\mu+\frac{1}{V}\sum_{{\bm k}'}(2U_0-U_{|{\bm k}-{\bm k}'|})n_{{\bm k}'}
\notag \\
 &\, +\frac{1}{(1-2\eta_{\bm k})V^2}
\sum_{{\bm k}_2{\bm k}_3{\bm k}_4}
\frac{\delta_{{\bm k}+{\bm k}_2+{\bm k}_3+{\bm k}_4,{\bm 0}}}{E_{{\bm k}}^{(0)}+E_{{\bm k}_2}^{(0)}+E_{{\bm k}_3}^{(0)}+E_{{\bm k}_4}^{(0)}}
\notag \\
&\,
\times U_{|{\bm k}+{\bm k}_2|}\biggl\{U_{|{\bm k}+{\bm k}_2|} (v_{{\bm k}_2}^2-u_{{\bm k}_2}^2)(u_{{\bm k}_3}v_{{\bm k}_4}+v_{{\bm k}_3}u_{{\bm k}_4})^2
\notag \\
& - U_{|{\bm k}+{\bm k}_3|}(v_{{\bm k}_2}v_{{\bm k}_3}-u_{{\bm k}_2}u_{{\bm k}_3})
(u_{{\bm k}_2}v_{{\bm k}_4}+v_{{\bm k}_2}u_{{\bm k}_4})
\notag \\
&\,\times (u_{{\bm k}_3}v_{{\bm k}_4}+v_{{\bm k}_3}u_{{\bm k}_4})\biggr\},
\label{xi}
\end{align}
\begin{align}
\Delta_{{\bm k}}= &\, -\frac{1}{V}\sum_{{\bm k}'}U_{|{\bm k}-{\bm k}'|}F_{{\bm k}'}
\notag \\
&\,  + \frac{2}{(1-2\eta_{\bm k})V^2}
\sum_{{\bm k}_2{\bm k}_3{\bm k}_4}
\frac{\delta_{{\bm k}+{\bm k}_2+{\bm k}_3+{\bm k}_4,{\bm 0}}}{E_{{\bm k}}^{(0)}+E_{{\bm k}_2}^{(0)}+E_{{\bm k}_3}^{(0)}+E_{{\bm k}_4}^{(0)}}
\notag \\
&\, 
\times U_{|{\bm k}+{\bm k}_2|}\biggl[ U_{|{\bm k}+{\bm k}_2|} u_{{\bm k}_2}v_{{\bm k}_2}(u_{{\bm k}_3}v_{{\bm k}_4}+v_{{\bm k}_3}u_{{\bm k}_4})^2
\notag \\
&\, - U_{|{\bm k}+{\bm k}_3|} u_{{\bm k}_2}v_{{\bm k}_3}
(u_{{\bm k}_2}v_{{\bm k}_4}+v_{{\bm k}_2}u_{{\bm k}_4})
\notag \\
&\,\times
 (u_{{\bm k}_3}v_{{\bm k}_4}+v_{{\bm k}_3}u_{{\bm k}_4})\biggr],
\label{Del}
\end{align}
where $E_{{\bm k}}^{(0)}$ is defined by
$E_{{\bm k}}^{(0)}\equiv (u_{{\bm k}}^2-v_{{\bm k}}^2)\xi_{{\bm k}}^{(0)}+2u_{{\bm k}}v_{{\bm k}}\Delta_{{\bm k}}^{(0)}$
in terms of $\xi_{{\bm k}}^{(0)}$ and $\Delta_{{\bm k}}^{(0)}$, which are obtained from  Eqs.\ (\ref{xi}) and (\ref{Del}) by omitting the correlation terms 
proportional to $V^{-2}$.
The solution of the equation for $\delta w_{\kappa_1\kappa_2\kappa_3\kappa_4}$, which is mentioned above, 
is also expressible in terms of $E_{{\bm k}}^{(0)}$ as
\begin{align}
&\,w_{{\bm k}_1\alpha_1{\bm k}_2\alpha_2{\bm k}_3\alpha_3{\bm k}_4\alpha_4}
\notag \\
=&\, -\frac{\delta_{{\bm k}_1+{\bm k}_2+{\bm k}_3+{\bm k}_4,{\bm 0}}}{E_{{\bm k}_1}^{(0)}+E_{{\bm k}_2}^{(0)}+E_{{\bm k}_3}^{(0)}+E_{{\bm k}_4}^{(0)}}
\frac{1}{V}
\biggl[\delta_{\alpha_1,-\alpha_2}\delta_{\alpha_3,-\alpha_4}
\notag \\
&\,\times (-1)^{1-\alpha_1-\alpha_3}
U_{|{\bm k}_1+{\bm k}_2|}(u_{{\bm k}_1}v_{{\bm k}_2}+v_{{\bm k}_1}u_{{\bm k}_2})
\notag \\
&\,\times (u_{{\bm k}_3}v_{{\bm k}_4}+v_{{\bm k}_3}u_{{\bm k}_4})
 +(\mbox{two terms})\biggr] ,
\label{w-def}
\end{align}
\end{subequations}
where $(\mbox{two terms})$ denotes terms obtained from the first term in the square brackets by 
the two cyclic permutations of $(2,3,4)$. 
This $w_{{\bm k}_1\alpha_1{\bm k}_2\alpha_2{\bm k}_3\alpha_3{\bm k}_4\alpha_4}$ is antisymmetric
in accordance with its original definition.

Equations (\ref{eta_k}), (\ref{nF-def}), (\ref{uv-sol}), and (\ref{xiDelta-def}) together with Eq.\ (\ref{N-constraint}) 
form closed nonlinear equations that can be used to evaluate the ground-state energy
of $s$-wave Cooper-pair condensation for any given potential $U(r)$.
Moreover, the corresponding normal state with correlations can be obtained by the replacement
\begin{align}
(u_{{\bm k}},v_{{\bm k}}) \longrightarrow (\theta(k-k_{\rm F}),\theta(k_{\rm F}-k)),
\label{uv-replace}
\end{align}
where $\theta(x)$ is the step function, and $k_{\rm F}$ is the Fermi wave number
at which $n_{\bm k}$ exhibits a discontinuity. 
Note that $k_{\rm F}$ remains invariant after switching on the interaction. \cite{Luttinger}
It should be noted that,
 in the limit of Eq.\ (\ref{uv-replace}) and $\eta_{\bm k}\rightarrow 0$,
Eq.\ (\ref{uv-replace}) reduces to the  normal ground-state energy evaluated by the
second-order perturbation expansion.

\subsection{Superposition over the number of Cooper pairs}

The operator $\hat\pi_4^\dagger$ in Eq.\ (\ref{|Psi>}) decreases the number of Cooper pairs 
by two, as seen from Eq.\ (\ref{pi_4}).
We thereby realize that $|\Phi_N\rangle$ is made up of a superposition over the number of Cooper pairs.
Indeed, the superposition can be quantified
by (i) expanding ${\cal B}_N$ of Eq.\ (\ref{calB_N})
in $(\hat\pi_4,\hat\pi_4^\dagger)$, (ii) sorting the series
in terms of the number of $(\hat\pi_4,\hat\pi_4^\dagger)$ pairs, 
and  (iii) multiplying the expansion by ${\cal B}_N^{-1}$ to normalize it.
The resultant probability $P_{\frac{N}{2}-2n}$ of having $\frac{N}{2}-2n$ Cooper pairs in the system is given,
within our approximation of retaining only the first term in the exponent of Eq.\ (\ref{calB_N}), by the Poisson distribution 
\begin{align}
P_{\frac{N}{2}-2n}=\frac{\lambda^n e^{-\lambda} }{n!}, \hspace{7mm} \lambda\equiv \frac{1}{2}\sum_{{\bm k}}\eta_{{\bm k}} ,
\label{P_n}
\end{align} 
where we have used  Eq.\ (\ref{eta_k}).
Note that $P_{\frac{N}{2}-2n}$ approaches a Gaussian distribution in the thermodynamic limit as seen from $\lambda\propto N$.

\section{Numerical Results}

\subsection{Model potential and numerical procedures}

Numerical calculations were performed for the model attractive potential
\begin{subequations}
\label{U-cal}
\begin{align}
U(r)\!=\!\frac{\hbar^2 a}{2mr_0^3}\,e^{-r/r_0}
\label{U(r)-cal}
\end{align}
with two parameters $a\!<\!0$ and $r_0\!>\!0$,
whose Fourier coefficients are given by
\begin{align}
U_q\!=\! \frac{4\pi\hbar^2 a}{m(1+r_0^2 q^2)}.
\label{U_q-cal}
\end{align}
\end{subequations}
The reason of using Eq.\ (\ref{U-cal}) with a finite range, 
instead of the contact attractive potential frequently used in the literature,
is to make our calculations free from the ultraviolet divergences inherent in the latter model.
Setting $(a,r_0)\!=\!(-0.12k_{\rm F}^{-1},0.1k_{\rm F}^{-1})$ yields 
a weak-coupling transition temperature $T_{\rm c}\!\approx\! 1.16\times 10^{-4}\varepsilon_{\rm F}^0/k_{\rm B}$, for example,\cite{KitaText}
where $\varepsilon_{\rm F}^0=\frac{\hbar^2k_{\rm F}^2}{2m}$ is the non-interacting Fermi energy and $k_{\rm B}$ is the Boltzmann constant.
Actually, we have chosen 
\begin{align}
(a,r_0)=(-0.19k_{\rm F}^{-1},0.1k_{\rm F}^{-1}),
\label{ar_0}
\end{align}
for which $T_{\rm c}\sim 2\times 10^{-2}\varepsilon_{\rm F}^0/k_{\rm B}$, 
to make the evaluation of the correlation parts in Eqs.\ (\ref{xi}) and (\ref{Del})
numerically tractable with high accuracy; Appendix B simplifies
the triple sums of wave vectors into triple radial and double angular integrals.
The radial integrals were performed over $0\leq k\leq k_{\rm cut}$ with cutoff $k_{\rm cut}\sim 50k_{\rm F}$ by expressing
$k=k_{\rm F}(1+\sinh x^3)$ and discretizing variable $x$ at an equal interval
so as to accumulate integration points around $k\sim k_{F}$.
It turned out that $E_{{\bm k}}^{(0)}$ defined below Eq.\ (\ref{Del}), which can be negative,
yields numerical instability when evaluating the quintuple integrals.
It was eventually removed by replacing every $E_{{\bm k}}^{(0)}$ in Eq.\ (\ref{xiDelta-def}) by the absolute value $|\xi_{{\bm k}}^{{\rm n}}|$ 
of the normal-state single-particle energy 
$\xi_{{\bm k}}^{{\rm n}}$ obtained from Eq.\ (\ref{xi}) by the replacement in Eq.\ (\ref{uv-replace}). 
The procedure corresponds to 
choosing $w_{\kappa_1\kappa_2\kappa_3\kappa_4}$ slightly away from the extremal value for numerical stability 
at the expense of increasing the variational ground-state energy.
Numerical calculations were performed by setting $\hbar=k_{\rm F}=2m=1$.
We have confirmed convergence with $\sim1\%$ error in the pair condensation energy
by choosing $k_{\rm cut}=50k_{\rm F}$ and having 130 (20) points for each radial (angular) integral.

\begin{figure}[t]
\begin{center}
\includegraphics[width=0.8\linewidth]{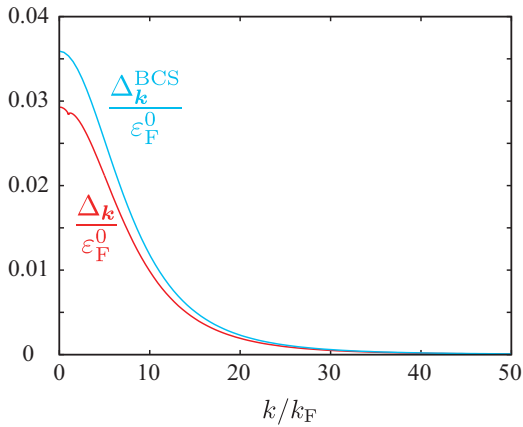}
\end{center}
\caption{(Color online) Energy gap $\Delta_{{\bm k}}$ in unit of $\varepsilon_{\rm F}^0=\frac{\hbar^2k_{\rm F}^2}{2m}$
as a function of $k/k_{\rm F}$ in comparison with $\Delta_{{\bm k}}^{\rm BCS}$ 
without the $\hat\pi_4$ correlations.
\label{Fig2}}
\end{figure}

\begin{table}[b]
\begin{center}
\begin{tabular}{c|cc}
 & $({\cal E}_{\rm n}-{\cal E}_0)/{\cal E}_0$ & $({\cal E}-{\cal E}_{\rm n})/{\cal E}_0$\\
\hline
Mean-field theory & $-6.877\times 10^{-2}$ & $-7.81\times 10^{-4}$ \\
With $\hat\pi_4$ correlations
& $-1.033\times 10^{-1}$ & $-5.06\times 10^{-4}$
\end{tabular}
\caption{Normal-state interaction energy ${\cal E}_{\rm n}\!-\!{\cal E}_0$ and pair condensation energy ${\cal E}\!-\!{\cal E}_{\rm n}$ in unit of 
the non-interacting kinetic energy ${\cal E}_0\!\equiv\! 2\sum_{{\bm k}}\varepsilon_k\theta(k_{\rm F}\!-\!k)$ for $(a,r_0)=(-0.19k_{\rm F}^{-1},0.1k_{\rm F}^{-1})$. \label{Table1}}
\end{center}
\end{table}

\subsection{Numerical results}

We present numerical results calculated for Eq.\ (\ref{ar_0}) self-consistently.
Figure \ref{Fig2} plots the energy gap $\Delta_{{\bm k}}$ as a function of $k/k_{\rm F}$ in comparison with $\Delta_{{\bm k}}^{\rm BCS}$
without the $\hat\pi_4$ correlations.
We observe that the correlations reduce the energy gap from the mean-field value and also 
produce a small dip around $k\!=\!k_{\rm F}$.
Table \ref{Table1} summarizes the corresponding ground-state energies.
As expected, the correlation energy due to $\hat\pi_4$  is seen to be much larger in magnitude than the pair condensation energy.
It should be noted that the mean-field condensation energy is still in excellent agreement with the BCS prediction \cite{BCS,Leggett06,SchriefferText}
\begin{align*}
-\frac{1}{2}N(0)(\Delta_{{\bm k}_{\rm F}}^{\rm BCS})^2\!=\!-7.81\times 10^{-4}{\cal E}_0
\end{align*}
given in terms of the density of states $N(0)\!=\!mk_{\rm F}V/2\pi^2\hbar^2$ and 
energy gap $\Delta_{{\bm k}_{\rm F}}^{\rm BCS}\!=\!0.0354\varepsilon_{\rm F}^0$ at the Fermi level.

\begin{figure}[t]
\begin{center}
\includegraphics[width=0.8\linewidth]{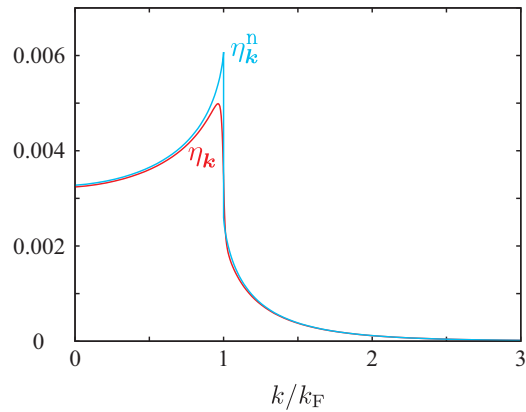}
\end{center}
\caption{(Color online) Plot of superconducting $\eta_{\bm k}$  
as a function of $k/k_{\rm F}$ in comparison with $\eta_{\bm k}^{\rm n}$ in the normal state.
\label{Fig3}}
\end{figure}

An important quantity that characterizes the correlations is $\eta_{{\bm k}}$ defined by Eq.\ (\ref{eta_k}).
In the normal state, it describes the deviation of Eq.\ (\ref{n_k}) from the non-interacting expression 
$n_{{\bm k}}^0\equiv \theta(k_{\rm F}-k)$ as  
\begin{align}
n_{{\bm k}}^{\rm n}=(1-\eta_{{\bm k}}^{\rm n})\,\theta(k_{\rm F}-k)+\eta_{{\bm k}}^{\rm n}\,\theta(k-k_{\rm F}),
\end{align}
and the resultant reduction of the discontinuity at $k=k_{\rm F}$ from 1. \cite{Luttinger}
Figure \ref{Fig3} shows $\eta_{\bm k}$ in the pair-condensed state in comparison with $\eta_{{\bm k}}^{\rm n}$
in the normal state. 
The latter exhibits a discontinuity of $\varDelta \eta_{{\bm k}}^{\rm n}=3.48\times 10^{-3}$ at $k=k_{\rm F}$, which
is blurred in $\eta_{{\bm k}}$ due to condensation.

\begin{figure}[t]
\begin{center}
\includegraphics[width=0.85\linewidth]{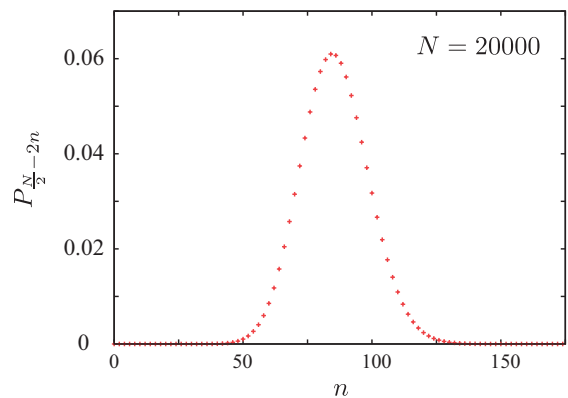}
\end{center}
\caption{(Color online) Probability $P_{\frac{N}{2}-2n}$ of having $\frac{N}{2}-2n$ Cooper pairs in the ket $|\Phi_N\rangle$
for $N=20000$.}
\label{Fig4}
\end{figure}

A finite $\eta_{{\bm k}}$ also produces a superposition over the number of Cooper pairs 
in the condensate that is expressed as Eq.\ (\ref{P_n}) in the weak-coupling region. 
Figure \ref{Fig4} shows the distribution of the number of Cooper pairs for $N=20000$,
which already has the appearance of a complete Gaussian.

\section{Concluding Remarks}

The present study has clarified that the correlations naturally produce a superposition over the number of Cooper pairs
in the ground-state wave function.
This superposition, which is given by Eq.\ (\ref{P_n}) and shown in Fig.\ \ref{Fig4},  enables us to define 
 the ``anomalous'' average unambiguously as Eq.\ (\ref{F_k}) within the number-conserving formalism,
in contrast to the mean-field BCS theory, where the average becomes finite only 
between states with different particle numbers as 
$\langle\Phi_{N-2}^{\rm BCS}|\hat{c}_{-{\bm k}\downarrow}\hat{c}_{{\bm k}\uparrow}|\Phi_{N}^{\rm BCS}\rangle$. \cite{Ambegaokar}
Indeed, the destruction of a single Cooper pair in our $|\Phi_N\rangle$ is accompanied by the creation of a pair of non-condensed particles.
Moreover, the gauge transformation $(\phi_{{\bm k}},w_{\kappa_1\kappa_2\kappa_3\kappa_4})\rightarrow (\phi_{{\bm k}}e^{2i\chi},w_{\kappa_1\kappa_2\kappa_3\kappa_4}e^{4i\chi})$ 
in Eqs.\ (\ref{|Phi>}) and (\ref{|Psi>}) changes Eq.\ (\ref{F_k}) as $F_{{\bm k}}\rightarrow F_{{\bm k}}e^{2i\chi}$
without affecting the ground-state energy. 
Thus, $F({\bm r}_1\!-\!{\bm r}_2)\equiv \sum_{{\bm k}}F_{{\bm k}}e^{i{\bm k}\cdot({\bm r}_1\!-\!{\bm r}_2)}$ has the property of 
a macroscopic wave function with a well-defined phase, which may vary from point to point in inhomogeneous systems.
It follows from Eq.\ (\ref{|Psi>}) that the superposition is
realized and sustained energetically by the exchange of quasiparticles between states with different numbers of Cooper pairs, 
similarly to the way that the coherence of two weakly coupled superconductors is sustained and mediated 
by the exchange of particles between them. \cite{Leggett06,Anderson66} 
Thus, the correlations are identified as being responsible for the emergence of macroscopic coherence in isolated superconductors.
The present study also makes it clear 
that {\em fluctuations in the number of condensed particles} $\varDelta N_{\rm con}$, instead of those in the total particle number as discussed frequently,
are responsible for the appearance of a macroscopic well-defined phase,
in accordance with the concept of off-diagonal long-range order based on reduced density matrices,\cite{Leggett06,Yang62}
the concept of coherence in optics, \cite{Glauber} and also the gauge invariance.

Thus, the present theory supports the mean-field description of superconductivity 
using the grand-canonical ensemble\cite{BCS,SchriefferText,Ambegaokar,KitaText}
in the thermodynamic limit.
For systems with a small number of particles or of low dimensions, on the other hand,
the fluctuations $\varDelta N_{\rm con}$ are expected to have substantial 
effects on the physical properties and realization of coherence. 
However, the present treatment cannot be applied directly to finite systems 
because of the approximation introduced around Eq.\ (\ref{P-def}), which becomes valid for $N\gg 1$.
We are planning to report some progress in removing the approximation in the near future.

\appendix

\section{Extremal Conditions\label{App:A_GA}}

The first variations of Eq.\ (\ref{barcalE}) with respect to $\phi_{{\bm k}}$ and $w_{\kappa_1\kappa_2\kappa_3\kappa_4}$ can be calculated concisely
with the chain rule.
Specifically, we introduce the following quantities in terms of the explicit dependences of $\bar{\cal E}$,
\begin{subequations}
\label{chain1}
\begin{align}
\xi_{{\bm k}}^{(0)}\equiv&\, \frac{1}{2}\frac{\delta\bar{\cal E}}{\delta n_{{\bm k}}}= \varepsilon_k-\mu+\frac{1}{V}\sum_{{\bm k}'}(2U_0 -U_{|{\bm k}-{\bm k}'|})  n_{{\bm k}'}
\label{chain1a}
\end{align}
\begin{align}
\Delta_{{\bm k}}^{(0)}\equiv  &\,-\frac{1}{2}\frac{\delta\bar{\cal E}}{\delta F_{{\bm k}}}= &\,-\frac{1}{V}\sum_{{\bm k}'}U_{|{\bm k}-{\bm k}'|} F_{{\bm k}'},
\label{chain1b}
\end{align}
\begin{align}
\frac{1}{2}\frac{\delta\bar{\cal E}}{\delta u_{{\bm k}}}=&\,\frac{1}{V}
\sum_{{\bm k}_2{\bm k}_3{\bm k}_4}\delta_{{\bm k}+{\bm k}_2+{\bm k}_3+{\bm k}_4,{\bm 0}}U_{|{\bm k}+{\bm k}_3|}
u_{{\bm k}_2}v_{{\bm k}_3}v_{{\bm k}_4}
\notag \\
&\,\times \sum_{\alpha\alpha'} (-1)^{1-\alpha-\alpha'}  w_{{\bm k}\alpha{\bm k}_2\alpha'{\bm k}_4-\alpha'{\bm k}_3-\alpha},
\end{align}
\begin{align}
\frac{1}{2}\frac{\delta\bar{\cal E}}{\delta v_{{\bm k}}}=&\, \frac{1}{V}
\sum_{{\bm k}_2{\bm k}_3{\bm k}_4}\delta_{{\bm k}+{\bm k}_2+{\bm k}_3+{\bm k}_4,{\bm 0}}U_{|{\bm k}+{\bm k}_3|}
 v_{{\bm k}_2}u_{{\bm k}_3}u_{{\bm k}_4}
\notag \\
&\,\times
\sum_{\alpha\alpha'} (-1)^{1+\alpha+\alpha'} w_{{\bm k}\alpha{\bm k}_2\alpha'{\bm k}_4-\alpha'{\bm k}_3-\alpha}.
\end{align}
\end{subequations}
Next, the derivatives of $(u_{{\bm k}},v_{{\bm k}},n_{{\bm k}},F_{{\bm k}})$ with respect to $\phi_{{\bm k}}$ 
can be calculated on the basis of Eqs.\ (\ref{uv}) and (\ref{nF-def}) as
\begin{subequations}
\label{chain2}
\begin{align}
\frac{\delta u_{{\bm k}}}{\delta \phi_{{\bm k}}}=  &\, -\frac{\phi_{{\bm k}}}{(1+\phi_{{\bm k}}^2)^{3/2}}=-u_{{\bm k}}^2v_{{\bm k}} ,
\\
\frac{\delta v_{{\bm k}}}{\delta \phi_{{\bm k}}}=&\, u_{{\bm k}}^3,
\\
\frac{\delta n_{{\bm k}}}{\delta \phi_{{\bm k}}} =&\,2\phi_{{\bm k}} u_{{\bm k}}^4 
(1-2\eta_{{\bm k}}),
\\
\frac{\delta F_{{\bm k}}}{\delta \phi_{{\bm k}}}=&\,(1-\phi_{{\bm k}}^2) u_{{\bm k}}^4(1-2\eta_{{\bm k}}) .
\end{align}
\end{subequations}
Similarly, the first variations of $(n_{{\bm k}},F_{{\bm k}})$ with respect to $w_{\kappa_1\kappa_2\kappa_3\kappa_4}$ 
are obtained from Eqs.\ (\ref{eta_k}) and (\ref{nF-def}), noting the comment below Eq.\ (\ref{eta_k}), as
\begin{subequations}
\label{chain3}
\begin{align}
\frac{\delta n_{{\bm k}}}{\delta w_{\kappa_1\kappa_2\kappa_3\kappa_4}}=&\,
\sum_{j=1}^4\delta_{{\bm k}_j{\bm k}}(u_{{\bm k}_j}^2-v_{{\bm k}_j}^2)w_{\kappa_1\kappa_2\kappa_3\kappa_4},
\\
\frac{\delta F_{{\bm k}}}{\delta w_{\kappa_1\kappa_2\kappa_3\kappa_4}}=&\,
-2\sum_{j=1}^4\delta_{{\bm k}_j{\bm k}}u_{{\bm k}_j}v_{{\bm k}_j} w_{\kappa_1\kappa_2\kappa_3\kappa_4}.
\end{align}
\end{subequations}
Using Eqs.\ (\ref{chain1}) and (\ref{chain2}), we can transform the extremal condition $\delta\bar{\cal E}/\delta\phi_{{\bm k}}=0$ into
\begin{align}
2\xi_{{\bm k}}^{(0)}\phi_{{\bm k}}+\Delta_{{\bm k}}^{(0)}(\phi_{{\bm k}}^2-1)+\chi_{{\bm k}} =0,
\label{phi_k-eq}
\end{align}
with
\begin{align}
\chi_{{\bm k}}\equiv &\, 
\frac{1}{V}
\sum_{{\bm k}_2{\bm k}_3{\bm k}_4}\delta_{{\bm k}+{\bm k}_2+{\bm k}_3+{\bm k}_4,{\bm 0}}U_{|{\bm k}+{\bm k}_3|}
\notag \\
&\,\times
\frac{u_{{\bm k}}v_{{\bm k}_2}u_{{\bm k}_3}u_{{\bm k}_4}-v_{{\bm k}}u_{{\bm k}_2}v_{{\bm k}_3}v_{{\bm k}_4}}{(1-2\eta_{{\bm k}})u_{{\bm k}}^2}
\notag \\
&\,
\times \sum_{\alpha\alpha'}(-1)^{1-\alpha-\alpha'} 
w_{{\bm k}\alpha{\bm k}_2\alpha'{\bm k}_3-\alpha'{\bm k}_4-\alpha},
\label{chi-def}
\end{align}
where we have used $(-1)^{\alpha+\alpha'}\!=\!(-1)^{-\alpha-\alpha'}$ for $\alpha,\alpha'\!=\!\pm\frac{1}{2}$.
Also using Eqs.\ (\ref{chain1}) and (\ref{chain3}), we can simplify $\delta\bar{\cal E}/\delta w_{\kappa_1\kappa_2\kappa_3\kappa_4}=0$ 
to 
\begin{align}
&\,2\sum_{j=1}^4\left[(u_{{\bm k}_j}^2-v_{{\bm k}_j}^2)\xi_{{\bm k}_j}^{(0)}+2u_{{\bm k}_j}v_{{\bm k}_j}\Delta_{{\bm k}_j}^{(0)}\right]w_{\kappa_1\kappa_2\kappa_3\kappa_4}
\notag \\
&\, +2\frac{\delta_{{\bm k}_1+{\bm k}_2+{\bm k}_3+{\bm k}_4,{\bm 0}}}{V}\biggl[U_{|{\bm k}_1+{\bm k}_2|}
(u_{{\bm k}_1}v_{{\bm k}_2}+v_{{\bm k}_1}u_{{\bm k}_2})
\notag \\
&\, \times (u_{{\bm k}_3}v_{{\bm k}_4}+v_{{\bm k}_3}u_{{\bm k}_4})\delta_{\alpha_1,-\alpha_2}\delta_{\alpha_3,-\alpha_4}(-1)^{1-\alpha_1-\alpha_3}
\notag \\
&\, +U_{|{\bm k}_1+{\bm k}_3|}
(u_{{\bm k}_1}v_{{\bm k}_3}+v_{{\bm k}_1}u_{{\bm k}_3})(u_{{\bm k}_4}v_{{\bm k}_2}+v_{{\bm k}_4}u_{{\bm k}_2})
\notag \\
&\, \times \delta_{\alpha_1,-\alpha_3}\delta_{\alpha_4,-\alpha_2}(-1)^{1-\alpha_1-\alpha_4}
\notag \\
&\, +U_{|{\bm k}_1+{\bm k}_4|}
(u_{{\bm k}_1}v_{{\bm k}_4}+v_{{\bm k}_1}u_{{\bm k}_4})(u_{{\bm k}_2}v_{{\bm k}_3}+v_{{\bm k}_2}u_{{\bm k}_3})
\notag \\
&\, \times \delta_{\alpha_1,-\alpha_4}\delta_{\alpha_2,-\alpha_3}(-1)^{1-\alpha_1-\alpha_2}\biggr]=0.
\label{w-eq}
\end{align}
In deriving the second term, we have performed tedious differentiations of the last term in Eq.\ (\ref{calE}) with respect to $w_{\kappa_1\kappa_2\kappa_3\kappa_4}$
and also used the identities $(-1)^{\alpha-\alpha'}\delta_{\alpha,-\alpha'}=-\delta_{\alpha,-\alpha'}$ and 
$(-1)^{\alpha+\alpha'}=(-1)^{-\alpha-\alpha'}$ for $\alpha,\alpha'=\pm \frac{1}{2}$.
Equation (\ref{w-eq}) can be solved formally to obtain Eq.\ (\ref{w-def}). 

Let us (i) substitute Eq.\ (\ref{w-def}) into Eq.\ (\ref{chi-def}), (ii) use 
\begin{align*}
&\,\sum_{\alpha\alpha'}(-1)^{2-2\alpha}\delta_{\alpha,-\alpha'}=\sum_{\alpha\alpha'}(-1)^{2-\alpha-\alpha'}\delta_{\alpha\alpha'}=-2,
\\
&\,\sum_{\alpha\alpha'}(-1)^{2-2\alpha-2\alpha'}=4,
\end{align*}
and (iii) exchange summation variables such as ${\bm k}_2\leftrightarrow{\bm k}_4$ several times.
We thereby find that Eq.\ (\ref{chi-def}) is expressible as
\begin{align}
\chi_{{\bm k}}= 2\xi_{\bm k}^{(1)}\phi_{{\bm k}}+\Delta_{\bm k}^{(1)}(\phi_{{\bm k}}^2-1) ,
\label{chi-2}
\end{align}
where $\xi_{\bm k}^{(1)}$ and  $\Delta_{\bm k}^{(1)}$ denote the correlation parts of Eqs.\ (\ref{xi})
 and (\ref{Del}), respectively, which are proportional to $V^{-2}$.
Substituting Eq.\ (\ref{chi-2}) into Eq.\ (\ref{phi_k-eq}), we obtain the equation for $\phi_{{\bf k}}$ as 
\begin{align}
2\xi_{\bm k}\phi_{{\bm k}}+\Delta_{\bm k}(\phi_{{\bm k}}^2-1) =0
\end{align}
in terms of Eqs.\ (\ref{xi}) and (\ref{Del}). 
The solution of this equation that satisfies $\phi_{{\bm k}}\rightarrow0$ for $k\rightarrow\infty$
is given by Eq.\ (\ref{phi_k}). 
 
\section{Sums over $({\bm k}_2,{\bm k}_3,{\bm k}_4)$}

Here we describe how to perform the triple sums
\begin{align}
f(k)\equiv&\,\frac{1}{V^2}\sum_{{\bm k}_2{\bm k}_3{\bm k}_4}\delta_{{\bm k}+{\bm k}_2+{\bm k}_3+{\bm k}_4,{\bm 0}}
\notag \\
&\,\times
U_{|{\bm k}+{\bm k}_2|}U_{|{\bm k}+{\bm k}_3|}g(k,k_2,k_3,k_4)
\label{f(k)}
\end{align}
efficiently, which is necessary for calculating Eqs.\ (\ref{xi}) and (\ref{Del}) numerically.
First, we choose ${\bm k}$ along the $z$ axis and express ${\bm k}_2$ in polar coordinates.
Then ${\bm k}+{\bm k}_2$ can be written as
\begin{align}
{\bm k}+{\bm k}_2=&\,(k_2\sin\theta_2\cos\varphi_2,k_2\sin\theta_2\sin\varphi_2,k+k_2\cos\theta_2)
\notag \\
= &\,(k_{12}\sin\theta_{12}\cos\varphi_2,k_{12}\sin\theta_{12}\sin\varphi_2,k_{12}\cos\theta_{12} ),
\label{k+k_2}
\end{align}
with
\begin{subequations}
\label{k_12thea_12}
\begin{align}
k_{12}\equiv&\, |{\bm k}+{\bm k}_2|=\sqrt{k^2+k_2^2+2kk_2\cos\theta_2},
\label{k_12}
\\
\theta_{12} \equiv&\, \arctan\frac{k_2\sin\theta_2}{k+k_2\cos\theta_2} .
\label{theta_12}
\end{align}
\end{subequations}
Equation (\ref{k+k_2}) is alternatively expressible in terms of the orthogonal matrix
\begin{align}
R_{12}\equiv \begin{bmatrix} \cos\theta_{12}\cos\varphi_2 & -\sin\varphi_2 & \sin\theta_{12}\cos\varphi_2 \\
\cos\theta_{12}\sin\varphi_2 & \cos\varphi_2 & \sin\theta_{12}\sin\varphi_2 \\
- \sin\theta_{12} & 0 &  \cos\theta_{12}
\end{bmatrix}
\end{align}
as
\begin{align}
{\bm k}+{\bm k}_2=R_{12}\begin{bmatrix} 0\\ 0\\ k_{12}\end{bmatrix} .
\end{align}
We also write ${\bm k}_3$ using $R_{12}$ as
\begin{align}
{\bm k}_3 =&\, R_{12}\begin{bmatrix} k_3\sin\bar\theta_3\cos\bar\varphi_3 \\ k_3\sin\bar\theta_3\sin\bar\varphi_3\\ k_3\cos\bar\theta_3\end{bmatrix}
,
\end{align}
where $(\bar\theta_3,\bar\varphi_3)$ are polar angles in the coordinate system where ${\bm k}+{\bm k}_2$ lies along the $z$ axis.
This representation enables us to write $({\bm k}+{\bm k}_2)\cdot{\bm k}_3$ and $|{\bm k}+{\bm k}_3|$ concisely as
\begin{align}
({\bm k}+{\bm k}_2)\cdot{\bm k}_3 
= k_{12}k_3 \cos\bar\theta_3,
\end{align}
\begin{align}
k_{13}\equiv|{\bm k}+{\bm k}_3|=&\, \left[k^2+k_3^2 + 2kk_3 \left(- \sin\theta_{12}\sin\bar\theta_3\cos\bar\varphi_3 \right.\right.
\notag \\
&\,\left.\left.+\cos\theta_{12} \cos\bar\theta_3 \right)\right]^{1/2} .
\label{k_13}
\end{align}
We can thereby transform Eq.\ (\ref{f(k)}) into
\begin{align*}
f(k)=&\, \frac{1}{(2\pi)^6} \int_0^\infty dk_2 k_2^2 \int_0^\pi d\theta_2\sin\theta_2\int_0^{2\pi} d\varphi_2
\notag \\
&\,\times \int_0^\infty dk_3 k_3^2\int_0^\pi d\bar\theta_3\sin\bar\theta_3\int_0^{2\pi} d\bar\varphi_3
U_{k_{12}}U_{k_{13}}
\notag \\
&\,\times g\left(k,k_2,k_3,\sqrt{k_{12}^2+k_3^2+2 k_{12}k_3\cos\bar\theta_3}\right).
\end{align*}
Integration over $\varphi_2$ can be performed easily to yield $2\pi$.
Subsequently, we make a change of variables $\bar\theta_3\!\rightarrow\!
k_4\!\equiv\!\sqrt{k_{12}^2+k_3^2+2k_{12}k_3\cos\bar\theta_3}$, with which $d\bar\theta_3\sin\bar\theta_3=-k_4dk_4/k_{12}k_3$,
to express $f(k)$ as
\begin{align}
f(k)=&\,  \frac{1}{(2\pi)^5} \int_0^\infty dk_2 k_2 \int_0^\infty dk_3 k_3   \int_0^\pi d\theta_2 \frac{k_2\sin\theta_2}{k_{12}}U_{k_{12}} 
\notag \\
&\,\times \int_{|k_{12}-k_3|}^{k_{12}+k_3}  
dk_4 k_4 g(k,k_3,k_3,k_4) 
\int_0^{2\pi} d\bar\varphi_3 U_{k_{13}} .
\label{f(k_1)-0}
\end{align}
Further, we exchange the order of integrations over $\theta_2$ and $k_4$ by noting that
$|k_{12}-k_3|\leq k_4\leq k_{12}+k_3$ is equivalent to $|k_3-k_4|\leq k_{12}\leq k_3+k_4$
and transforming the latter into
\begin{align}
\frac{(k_3-k_4)^2-k^2-k_2^2}{2kk_2} \leq \cos\theta_2 \leq \frac{(k_3+k_4)^2-k^2-k_2^2}{2kk_2}.
\label{costheta_2}
\end{align}
The two inequalities are satisfied when $\frac{(k_3-k_4)^2-k^2-k_2^2}{2kk_2}\leq 1$ and 
$\frac{(k_3+k_4)^2-k^2-k_2^2}{2kk_2}\geq -1$ are simultaneously met, which are transformed into $k_{4{\rm i}}\leq k_4\leq k_{4{\rm f}}$ with
\begin{subequations}
\begin{align}
\left\{\begin{array}{ll} k_{4{\rm i}}\equiv{\rm max}\left(0,k_3-k-k_2,|k-k_2|-k_3\right)\\
k_{4{\rm f}}\equiv k+k_2+k_3\end{array}\right. .
\label{k_4-condition}
\end{align}
In addition, Eq.\ (\ref{costheta_2}) is expressible in terms of  two angles $(\theta_{2{\rm i}},\theta_{2{\rm f}})$ defined through
\begin{align}
\left\{\begin{array}{ll} \displaystyle
\cos\theta_{2{\rm i}}\equiv {\rm min}\left(1,\frac{(k_{4}+k_3)^2-k^2-k_2^2}{2kk_2}\right)\\
\displaystyle\cos\theta_{2{\rm f}}\equiv {\rm max}\left(-1,\frac{(k_{4}-k_3)^2-k^2-k_2^2}{2kk_2}\right)
\end{array}\right. 
\end{align}
\end{subequations}
as $\theta_{2{\rm i}}\leq \theta_2\leq \theta_{2{\rm f}}$.
We can thereby transform Eq.\ (\ref{f(k_1)-0}) into
\begin{align}
f(k)=&\,  \frac{1}{(2\pi)^5} \int_0^\infty dk_2 k_2  \int_0^\infty dk_3 k_3  \int_{k_{{4\rm i}}}^{k_{{4\rm f}}}  dk_4 k_4  
\notag \\
&\,\times g(k,k_2,k_3,k_4)
\int_{\theta_{2{\rm i}}}^{\theta_{2{\rm f}}} d\theta_2\sin\theta_2 \frac{k_2}{k_{12}}U_{k_{12}} 
\notag \\
&\,\times \int_0^{2\pi} d\bar\varphi_3 U_{k_{13}} ,
 \label{f(k)-final}
\end{align}
where $k_{12}$ and $k_{13}$ are respectively defined by Eqs.\ (\ref{k_12}) and (\ref{k_13}) in terms of $\theta_{12}$ and $\bar\theta_3$ given by Eq.\ (\ref{theta_12})
and  
\begin{align}
\bar\theta_3\equiv &\,\cos^{-1}\frac{k_4^2-k_{12}^2-k_3^2}{2k_{12}k_3} .
\end{align}
The last integral over $\bar\varphi_3$ in Eq.\ (\ref{f(k)-final}) can be performed analytically
for the present model given by Eq.\ (\ref{U_q-cal})
as may be seen  from Eq.\ (\ref{k_13}).


\begin{thebibliography}{99}
\bibitem{BCS}J. Bardeen, L. N. Cooper, and J. R. Schrieffer, Phys. Rev. {\bf 108}, 1175 (1957).
\bibitem{Schrieffer11}J. R. Schrieffer, in {\em BCS: 50 Years}, ed. L. N. Cooper and D. Feldman (World Scientific, Singapore, 2011) p. 21.
\bibitem{Peierls}R. Peierls, Contemp. Phys. {\bf 33}, 221 (1992).
\bibitem{Leggett06}A. J. Leggett, {\em Quantum Liquids: Bose Condensation and Cooper-Pairing in Condensed Matter Systems} (Oxford Univ. Press, Oxford, 2006).
\bibitem{Anderson66}P. W. Anderson, Rev. Mod. Phys. {\bf 38}, 298 (1966).
\bibitem{Anderson64}P. W. Anderson, in {\em The Many-Body Problem}, ed. E. R. Caianiello (Academic, New York, 1964)  Vol. 2, p. 113.
\bibitem{Leggett91}A. J. Leggett and F. Sols., Found. Phys. {\bf 21}, 353 (1991).
\bibitem{SchriefferText}J. R. Schrieffer, {\em Theory of Superconductivity} (W.A. Benjamin, Reading, 1964) p. 48.
\bibitem{Ambegaokar}V. Ambegaokar, in {\em Superconductivity}, ed. R. D. Parks (Marcel Dekker, New York, 1969)  Vol. 1, Chap.\ 5, Sect. IIC.
\bibitem{Kita17}T. Kita, J. Phys. Soc. Jpn. {\bf 86}, 044003 (2017).
\bibitem{JY96}J. Javanainen and S. M. Yoo, Phys. Rev. Lett. {\bf 76}, 161 (1996).
\bibitem{CD97}Y. Castin and J. Dalibard, Phys. Rev. A {\bf 55}, 4330 (1997).
\bibitem{Leggett80}A. J. Leggett, in {\em Modern Trends in the Theory of Condensed Matter}, ed. A. P\c{e}kalski and J. Przystawa (Springer-Verlag, Berlin, 1980), p. 13.
\bibitem{NSR85}P. Nozi\`eres and S. Schmitt-Rink, J. Low Temp. Phys. {\bf 59}, 195 (1985).
\bibitem{KitaText}T. Kita, {\em Statistical Mechanics of Superconductivity} (Springer, Tokyo, 2015) Sect.\ 9.2.
\bibitem{Luttinger}J. M. Luttinger, Phys. Rev. {\bf 119}, 1153 (1960).
\bibitem{Yang62}C. N. Yang, Rev. Mod. Phys. {\bf 34}, 694 (1962).
\bibitem{Glauber}R. J. Glauber, Phys. Rev. {\bf 131}, 2766 (1963).


\end{thebibliography}
\end{document}